\documentclass[conference]{IEEEtran}

\usepackage{balance}

\IEEEoverridecommandlockouts 
\usepackage{graphicx}
\usepackage{algorithmic}
\usepackage{algorithm}
\usepackage{listings}
\usepackage[OT4,T1]{fontenc}
\usepackage[cmex10]{amsmath}
\usepackage{url}
\usepackage{multirow}

\usepackage{dirtytalk}

\usepackage[backend=biber]{biblatex}
\title{MyMigrationBot: A Cloud-based Facebook Social Chatbot for Migrant Populations}
\author{
\IEEEauthorblockN{Karol Chlasta, Paweł Sochaczewski, Izabela Grabowska, Agata Jastrzębowska}
\IEEEauthorblockA{
Kozminski University\\
CRASH Center for Research on Social Change and Human Mobility\\
ul.\ Jagiellońska 57/59, 03-301 Warsaw, Poland\\
Email: kchlasta@kozminski.edu.pl}
}

\begin{document}
\maketitle              

\begin{abstract}
We present the design, implementation and evaluation of a new cloud-based social chatbot called MyMigrationBot, that is deployed to Facebook. The system asks and answers questions related to user's personality traits and person-job competency fit to give feedback, and potentially support migrant populations. The chatbot's response database is based on reputable socio-psychological tools and can be customised. The system's backend is written with Node.js, deployed to AWS and Twilio, and joined with Facebook through Graph and Messenger APIs. To our knowledge this is the first multilingual social chatbot deployed to Facebook and designed to research and support migrant populations with feedback in Europe. It does not have personality like other bots, but it can study and feedback on migrants' personality and on other customised questionnaires e.g., job-competency fit. The aim of a social chatbot in our research project is to help engage migrants with social research using feedback information tailored to them. It can help migrants to get knowledge about their psycho-social resources and therefore to facilitate their integration process into a receiving labour market. We evaluated the chatbot on a group of 53 people, incl. 23 migrants, and we present the results.
\end{abstract}

\section{Introduction}
\IEEEoverridecommandlockouts\IEEEPARstart{C}{hatbots} are increasingly popular in everyday interactions. The ever increasing affordability and popularity of ICT devices leads to a wider range of communication channels available for different social groups. Is might be viable to use these channels for a number of reasons. Firstly, a social chatbot as an interaction system is able to gather feedback from its respondents and use it to increase the availability of information, the quality of both commercial and public services, and perhaps improve users' quality of life. Secondly, a social chatbot can give personalised feedback to respondents which can increase motivation to participate in research. And thirdly, a social chatbot with feedback can give higher satisfaction out of participating in research~\cite{kuhne2018personalized}.

This manuscript focuses on \emph{MyMigrationBot}, a social chatbot system able to gather Big Data for social research from Facebook, and interact with users through multiple communication channels using a Facebook Messenger protocol.

\section{Motivation}
Migrant populations often lack support, usually from available public services in a receiving country, as there are massive movements in a short period of time, due to some unpredictable circumstances like war or a natural disaster. Such people are usually not prepared to start a new working life in a new labour market.

In recent years Poland, as well as other European countries received millions of both labour immigrants and war refugees. We believe these people could be supported in the receiving societies not only by human agents, but also by new technologies like personal assistants and avatars, often simply known as the social chatbots.

In 2022 a new massive wave of immigration from Ukraine arrived in Poland. As of the 5th of May 2022 according to Operational Data Portal of UN High Commissioner for Refugees\footnote{https://data2.unhcr.org/en/countries/} 3,143,550 Ukrainians entered Poland out of 5,757,014 people who left Ukraine since February 2022, when the Russian invasion had started, which consists of nearly 55 per cent of all recently fleeting Ukrainian population.

Immigrants and refugees widely use mobile phones and other smart devices. The poly-media accessibility in one smartphone makes it 'a must-to-be taken' by every human on the move~\cite{madianou2014smartphones}. As Dekker et al. (2018)~\cite{dekker2018smart} show in relation to Syrian refugees who applied for asylum in the European Union (EU) member states in 2015 and 2016 (the largest group), they used smartphones and social media in migration decision making. The most meaningful issues for refugees as reported by the authors are linked to the access and evaluation of the trustworthiness of information. The kind of social chatbot which we propose in this article could for instance help them to diagnose their psycho-social capitals and to facilitate navigation through the available public services upon arrival.

This is because our chatbot can use different structured or semi-structured questionnaires in its conversations. Apart from that, new custom conversations can be designed to enhance the existing functionality of the~\emph{MyMigrationBot}. One of the biggest advantages of our chatbot is the functionality of feedback given to our users as a 'social and non-material' complimentary thank you for their participation. The feedback can be also customised.  

To summarise, the main motivation for us was to create a multilingual social chatbot system - as a conversational agent - able to provide migrant populations with targeted, immediate feedback, so that they can better integrate into a receiving society.

\section{Questions and Feedback}
The first set of questions asked by \emph{MyMigrationBot} is taken from the Ten-Item Personality Inventory (TIPI)~\cite{gosling2003very} measuring the Big Five personality dimensions: (1) Extroversion, (2) Agreeableness, (3) Conscientiousness, (4) Emotional Stability, and (5) Openness to Experience. As the original TIPI questionnaire was created in English only, we used the Polish adaptation by Sorokowska (2014)~\cite{sorokowska2014polska}, and a translation to Ukrainian prepared by a native speaker of Ukrainian. Kosinski at al. (2014)~\cite{kosinski2014manifestations} show manifestations of user personality in website choice and behaviour on online social networks, which was a great inspiration for us. 

Personality has been linked to migration for a number of reasons. The work offered by social psychologists goes a long way to show that the migrant personality matters in making migration decisions. Personality traits influence international voluntary migration. They also inform about self-selection to migration. Boneva and Frieze (2001)~\cite{boneva2001toward} show that individuals who want to emigrate possess a syndrome of personality characteristics that differentiates them from those who want to stay in their country of origin. They also explain the role of personality in desires to emigrate. Emigrants are not just responding to a particular set of economic conditions; there is something specific about the personality of those who desire to move. Emigrants are less prone to anxiety and insecurity than non-emigrants~\cite{ray1986traits}. Higher persistence, openness to experience, as well as previous experience of living internationally, all increased the chances that a participant was planning to move abroad. Higher agreeableness and conscientiousness lowered the odds of a move. Men who were lower in emotional stability were more likely to want to leave, but the same effect was not found for women~\cite{tabor2015migrant}. Liable et al. (2021)~\cite{laible2021does} showed that non-cognitive personality traits explain the wage gap between male migrants and non-migrants. Polek et al. (2011)~\cite{polek2011evidence} report that personality traits bring us information about adjustments of migrants to the new environments, and they can help in migrant integration processes.

Yet another tool that we incorporated into our \emph{MyMigrationBot} was a Job-Competency Fit Scale~\cite{jastrzkebowska2020dopasowanie}, which also gives tailor-made feedback to our respondents. This measure of fit is referred to in psychology as a molar measure, a direct question~\cite{edwards2006phenomenology}. Although a competence has many meanings, the main meaning is about performing a task, or a function on an adequate level, with knowledge and understanding of a situation. Job fit is the key factor in being hired today in the labour market. We endowed our social chatbot with the list of 26 human competencies connected to cognitive and soft/social competencies which were tested initially in big social surveys on human capital~\cite{grabowska2022migration}. Then we asked our users: "Think about the job you are currently working in. Then, please specify to what extent the competencies and skills listed below are needed (required) for the position you hold?". Then the chatbot gives feedback according to means verified before.
Therefore our social chatbot offers a tool which can help to match a person and their competencies to a proper job.

User experience is a key concept for designing and enhancing the quality and usability of software products~\cite{vermeeren2010user}. Therefore, we have decided to include System Usability Scale~\cite{bangor2008empirical} (SUS) to gather some subjective assessment on the \emph{MyMigrationBot}'s design and usability for migrants and non-migrants. It was a simple survey with a standard scale consisting of a 10 item questionnaire with five response options for respondents: from strongly agree to strongly disagree (Likert's social scale). We need to learn more about the bot's accuracy, task completion and responsiveness. User experience evaluation allows us to identify how our MyMigrationBot should evolve in the future to meet experience and expectations~\cite{moreno2013hci} of our target population - various categories of migrants. To a certain degree, the assessment feedback by users embeds us into a co-designing process~\cite{chen2020creating}, especially since our MyMigrationBot is still in a prototype phase. 

Nowadays many chatbots have been designed with an aim to assist with information-seeking, and guidance are based on a frequently asked question and answers mode (FAQ). Sansonett et al. argue that human users expect chatbots to understand their texts, provide adequate answers and that they will be interactive with humans in run-time.
As far as we know there are many social chatbots to support psychotherapy (e.g. Woebot) and AI legal services. There are chatbots who are given personality (e.g. XiaoIce of Microsoft) in order to make them user-friendly and responsive to users' queries and interests. There is also a Tinka chatbot - a dialogue system performing as an assistant in the mobile phone delivery system of T-mobile Austria. Tinka is able to act and deliver customer information in various topics. 
Our benchmark might be Eike, a chatbot (with an avatar) with a personality designed to deliver information to various groups of migrants~\cite{chen2020creating}. According to respondents' testing this chatbot, Eike \say{should be be a gentle city-born messenger pursuing peace in the neighbourhood. Eike should be able to know about living in a German city and be happy to share their knowledge with anyone who comes to seek it. Eike should soothe worries in a soft and friendly voice and always have a positive appearance, rendering migrants hopeful and optimistic in terms of living and working} (2020: 224). Our approach presented in this article is different. We design, implement and evaluate a social chatbot called \emph{MyMigrationBot} which helps users to learn more about their personality (a bot does not have a personality of its own), their Job-Competency Fit and possible other customised questionnaires with feedback diagnosing psycho-social capital of humans. Therefore our conversation design is slightly different than 'a standard' chatbot design. Next to introduction, greetings, admitting errors, it delivers answers. The bot is particularly designed for migrants to help them to diagnose themselves for the labour market's needs. 

\section{Implementation}
The architecture of \emph{MyMigrationBot} comprises of front-end in Facebook Messenger, and back-end deployed into AWS Cloud. The conversation engine uses Twilio~\cite{janarthanam2017hands} platform, that is linked to Facebook using Facebook Send API\footnote{https://developers.facebook.com/docs/messenger-platform}. All data is stored with AWS RDS service with MySQL 8 engine. The source code of the \emph{MyMigrationBot} is stored in a private GitHub repository\footnote{https://github.com/KarolChlasta/BigMig}, shareable with interested parties on-demand. Back-end is written with Node.js version 16.13.1, and it uses Crypto package for Facebook application secrets encryption. We use Node.js MySQL drivers (version 2.18.1) for accessing the database\footnote{https://www.npmjs.com/package/mysql}, and the Node.js request package\footnote{https://www.npmjs.com/package/request} (version 2.88.2) for processing https requests to Facebook's Graph API endpoint (in version 12). Node.js Twilio package\footnote{https://www.npmjs.com/package/twilio} (version 3.76.1) is used to invoke calls to Twilio Autopilot API endpoint. Both Twilio Autopilot and Twilio Functions were created manually using the Twilio console. We also used Node.js Twilio-cli package\footnote{https://twil.io/cli} to manage our Twilio resources (e.g. Autopilot and Functions).

\begin{figure*}[tbp]
\centering
\includegraphics[width=1\hsize]{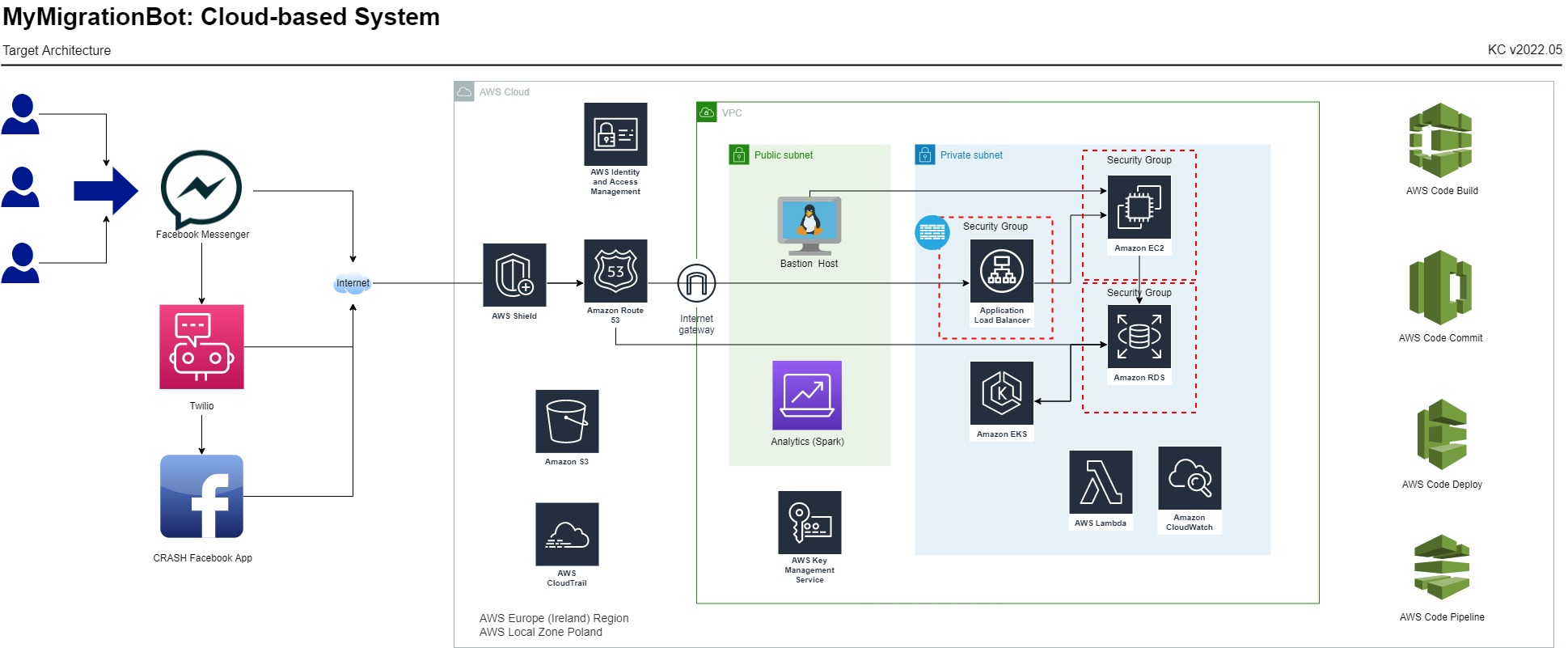}
\caption{Target Architecture of MyMigrationBot}
\label{TargetArchitecture}
\end{figure*}

\subsection{Dialogue Management}
To drive the responses of \emph{MyMigrationBot} we used Twilio Autopilot, a response and dialogue management system based on customizable, serverless functions from the Twilio cloud platform. The Autopilot service~\cite{firth2019new} allows us to \say{build, train, and deploy AI bots using natural language understanding and machine learning}. It also simplifies deployment of the required functions from the code repository to different cloud-based environments.

The dialogue manager functionality within Twilio Autopilot chooses which response to return to the user. The system will choose "the best response", so the one that was ranked highest by the engine. If the score of the top ranked response is below a defined threshold (determined and customizable in the Autopilot's configuration), the dialogue manager system will select an off-topic response instead, that indicates lack of understanding (e.g. say “I do not understand, please repeat.”). The system also contains a simulator allowing to both test and train the chatbot.

Apart from Twilio, the \emph{MyMigrationBot} conversations can be monitored through the administration panel of a Facebook Page, to which it is linked. Moreover, a Facebook user granted page administrator role can monitor and participate in chatbot's conversations. As a result a human agent is able help users facings issues in conversations, to help them progress through the survey, or to help the chatbot in responding to any non-standard interactions, e.g. to avoid biased contents of databases~\cite{schlesinger2018let}.

While Twilio platform has recently been used for custom chat applications using WhatsUp by Immigration Policy Lab at Stanford University~\cite{fei2020automated}, this is the first deployment of Twilio cloud platform with Facebook Messenger to migration studies, and second to support migrants in Europe.

\subsection{Facebook APIs}
Facebook launched the Messenger platform in 2016\footnote{https://about.fb.com/news/2016/04/messenger-platform-at-f8/}. We use Facebook Messenger API through Twilio platform. The API connects to the backend of \emph{MyMigrationBot}, which was deployed to AWS Cloud, and it bridges Twilio with Facebook. When a message event occurs, it notifies bot's web-hook and calls a predefined Twilio function.

We also use Facebook Graph API\footnote{https://developers.facebook.com/docs/graph-api/} in its latest version (v13.0). As stated
by Facebook, the Graph API is the primary way to
read and write to the Facebook social graph, and all their SDKs and products \say{interact with the Graph
API in some way}. We used this API to retrieve a
limited set of demographic information
about~\emph{MyMigrationBot}'s users from their
public Facebook profiles. We recognise the fact that due to recent controversies and data leaks~\cite{mcmanaman2019strategic}, the use of Facebook's Graph API is now made more difficult for any non-public elements of the profile. The process now requires a multi-level approval, and a time-consuming vetting process from Meta, Facebook's parent company. This might impact the timelines of adding any new attributes to our database in future.

At the moment, each time a conversation with \emph{MyMigrationBot} is started, the basic set of attributes is saved into the RDS in AWS Public
Cloud. These attributes are described in Table~\ref{RDSDatabaseStructure} of the Appendix section.

Once all the back-end actions of \emph{MyMigrationBot} are complete, the relevant text message is selected using the dialogue management system, and the response is sent to Facebook to deliver to the users' Messenger client(s). The process ends with matching the user's responses to the variables for the relevant user session. A screenshot of multilingual interaction with \emph{MyMigrationBot} is presented in Figure~\ref{TargetArchitecture}.

\subsection{Infrastructure}
The infrastructure of this project is hosted in AWS Cloud. A few of the AWS services and their components were created, and configured manually (e.g. Virtual Private Cloud, Public and Private Sub-nets, Network ACLs and IAM users). Other AWS services, like RDS, EC2, Security Groups, Launch configuration, and Auto Scaling groups were deployed automatically using Terraform (version 1.0.11). The Infrastructure was loaded from the code by Terraform using development workstations. Terraform state files are kept in Amazon S3. Terraform gets access to AWS Cloud via AWS IAM user access keys. The same method of authentication is used by AWS Command Line Interface to manage AWS resources for the project.

Our infrastructure in AWS is kept in Virtual Private Cloud (VPC) in Europe (Ireland) region. We used Internet Gateway Component to open network traffic between the public Internet and our VPC.

We protect our infrastructure with two layers of firewall, on the subnet and application level. We use Network ACLs for subnet level firewall and Application Firewall for Security Groups, for which we opened the traffic only for the protocols and ports that are used in the project. The Back0end was installed on a single EC2 instance, which was created via Auto Scaling Groups. The Server Instance Type of t2.micro (1 CPU, 1 GiB Memory, 3.3 GHz Intel Scalable Processor) was selected for the beta testing.

To protect our back-end server's webhooks against execution from unauthorised third-party applications we use Facebook Page and Verify Tokens, as well as application secrets.

Apart from having a specific domain name hosted in AWS Route 53, we also used Ngrok service to publish URL for our back-end server. We haven’t used the AWS Application Load Balancer, as it was not justified by the beta testing stage of the project.

The back-end server is run as a Windows service. It is started automatically during the boot process of the EC2 instance. 

\section{Evaluation}
\subsection{Participants}
We recruited N=53 participants for the study. Mostly from Poland (n = 34; 64.2\%), men (n = 33; 62.26), filling the tool on the computer (n = 25; 47.2\%) or on the phone (n = 20; 37.7\%). Almost half of them are people with experience of migration (n = 23; 43.4\%). Mean age (M = 31.02; SD = 12.73). Medium ICT skills (M = 4.15 / 5.00; SD = 0.91). Note that the male group includes a single person, who declares a gender that is not aligned with the person's legal status in the country. Although every participant was an active Facebook user, only 21 (39.6\%) declared their ICT skills as excellent. Detailed data on participants is presented in Table~\ref{EvalDemo}.

\begin{table}[tbp]
\caption{Descriptive characteristics of the participants}\label{EvalDemo}
\centering
\begin{tabular}{|@{\vrule width0ptheight9pt\enspace}l|c|c|c|}\hline
\hfil\bf Variable Name&\bf Category&\bf Number& \bf\%\\\hline
\multirow{8}{*}{Nationality}&Polish&34&64.2\\\cline{2-4}
&Ukrainian&6&11.3\\\cline{2-4}
&Belarusian&5&9.4\\\cline{2-4}
&British&1&1.9\\\cline{2-4}
&Czech&1&1.9\\\cline{2-4}
&French&1&1.9\\\cline{2-4}
&Polish-Italian&1&1.9\\\cline{2-4}
&No data&4&7.5\\\hline
\multirow{8}{*}{Country of Birth}&Polish&34&64.2\\\cline{2-4}
&Ukraine&6&11.3\\\cline{2-4}
&Belarus&5&9.4\\\cline{2-4}
&UK&2&3.8\\\cline{2-4}
&Czechia&1&1.9\\\cline{2-4}
&France&1&1.9\\\cline{2-4}
&India&1&1.9\\\cline{2-4}
&No data&3&5.7\\\hline
\multirow{2}{*}{Gender}&Female&20&37.73\\\cline{2-4}
&Male&32&62.26\\\hline
\multirow{3}{*}{Device}&Computer&25&47.2\\\cline{2-4}
&Mobile Phone&21&39.6\\\cline{2-4}
&No data&7&13.2\\\hline
\multirow{2}{*}{Immigrant}&No&30&56.6\\\cline{2-4}
&Yes&23&43.4\\\hline
\end{tabular}
\end{table}

\subsection{Procedure}
\begin{figure*}[tbp]
\centering
\includegraphics[width=1\hsize]{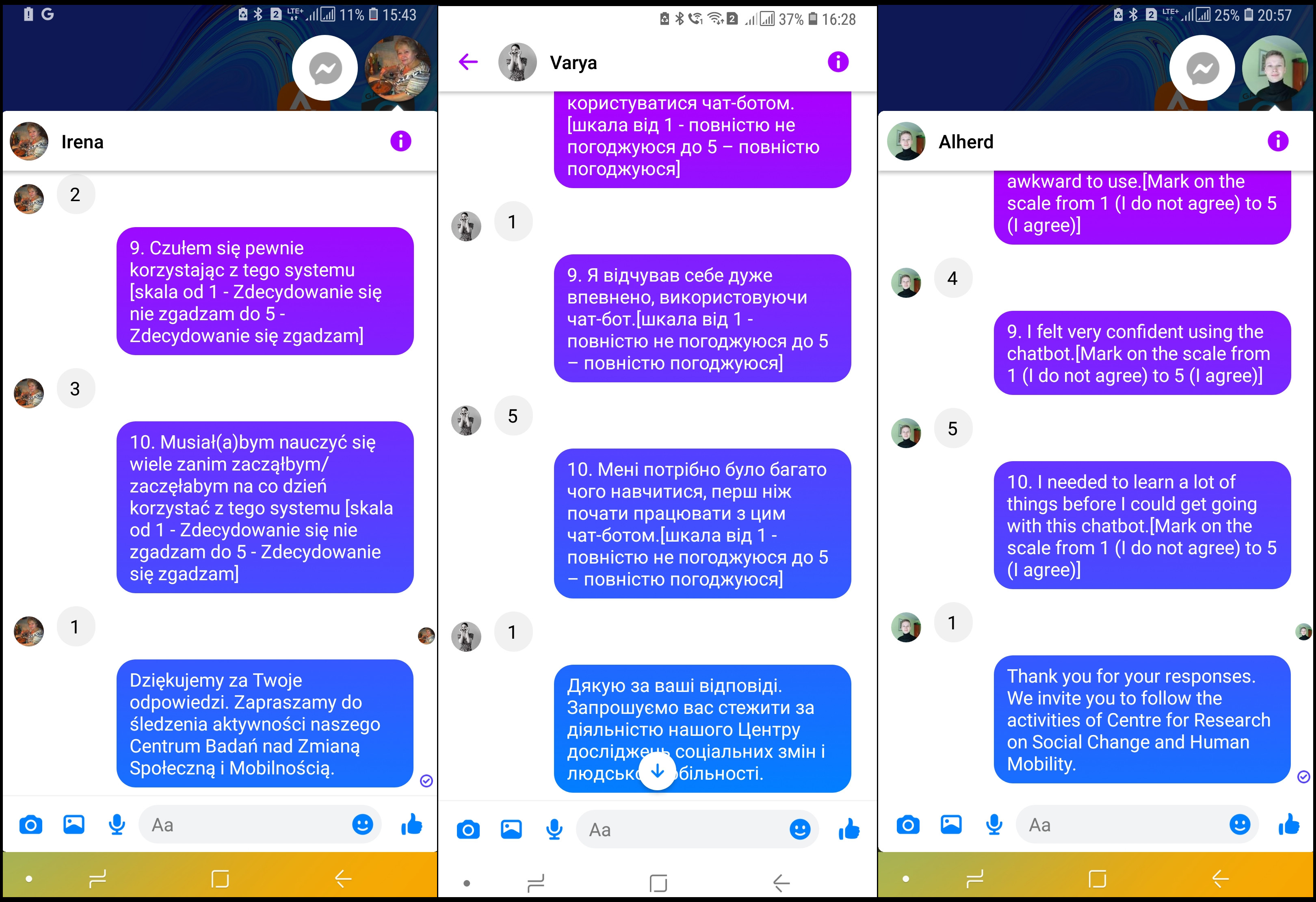}
\caption{\emph{MyMigrationBot}'s interaction with users of mobile devices on Facebook Messenger in Polish, Ukrainian and English}
\end{figure*}

The participants were engaged with a pilot study of MyMigrationBot via a Facebook page of CRASH Center for Research on Social Change and Human Mobility of Kozminski University using Facebook Messenger\footnote{https://www.facebook.com/CRASHKozUni}.

Participants used their own personal computers or mobile devices to interact with \emph{MyMigrationBot}. After completing a single interaction with the chatbot, they were asked to answer a post-study questionnaire on the usability of that interaction, with optional open-ended questions on \emph{MyMigrationBot}'s advantages and disadvantages. In took around 5-10 minutes to complete the study.

\section{Results}
System Usability Scale (SUS) was used. This is a 10 item questionnaire with five response options for respondents; from Strongly agree to Strongly disagree. SUS allows us to evaluate a wide variety of products and services, including hardware, software, mobile devices, websites and applications~\cite{bangor2008empirical}. Though the scores are 0-100, SUS score above 68 would be considered above average and anything below 68 is below average.

Both on the computer and on the phone, the \emph{MyMigrationBot} respondents' results at SUS were in both cases about $M = 70.00$ (computer $M = 70.00$ and mobile phone $M = 70.25$).

To compare the System Usability Scale between respondents who experienced and did not experience migration, t-Student for independent groups analysis was counted. It turned out that people with no experience of migration assessed the tool usability slightly better ($M = 71.08; SD = 8.14$) than people with experience of migration ($M = 68.26; SD = 12.14$). However, this difference turned out to be statistically insignificant $t(51) = 1.012; p> 0.05$.

We have also gathered additional, optional, unstructured feedback from the participants in Table~\ref{ParticipantFeedback} of the Appendix section of this manuscript. The feedback was based on 5 open-ended questions.
\section{Discussion}
With the rapid inflow of migrants to Europe the social chatbots have their momentum~\cite{chen2020creating} both for research, and for practical use associated with information-seeking and migrant integration as a result.

Our \emph{MyMigrationBot} is not designed as a persona with a personality but as a professional trustworthy assistant to diagnose a respondent's personality and other psycho-social resources through an individual tailor-made feedback protocol. The architecture of our bot is flexible enough to customise questionnaires with feedback when new needs emerge. The main limitation is that we have not been able to fully implement the target architecture in relation to all the AWS cloud components and automation services. As a result, the maintenance of the \emph{MyMigrationBot} is still partially manual. The configuration of the server infrastructure (AWS EC2, Ngrok) was automated using the user data script that gets executed on the instance’s initial boot. The script downloads and installs all the required utilities e.g. AWS Client, Ngrok, as well as retrieving the source code from GitHub, but that script is still triggered from the developers' laptops.

At present we also do not use all the services listed on Figure~\ref{TargetArchitecture}, namely the Analytics component (Spark), as well as the AWS CodePiplines, CodeBuild, and CodeDeploy. We use a free GitHub repository instead. The reason for that limitation was cost. Our intention was to cut the cost of the project until the beta testing finishes. All the cloud infrastructure is attached to the main author's individual Twilio and AWS accounts (and his credit card). This was also the reason for a technical decision to host our Internet backend with Ngrok service.

The \emph{MyMigrationBot} was also tested on limited number of users, with the maximum of 5 users using the system at the same time on April 28th 2022. We recognise that more in depth performance testing is needed prior to the production release of the application. 

Additionally, manual configuration of Twilio Autopilot and Twilio Functions is prone to human error. In future, we want to be able to automate creation of Autopilot (together with Twilio Services, that replace Twilio Functions)  directly from the source code, using Twilio CLI. We believe this will simplify building new chatbots, and it will be convenient for multi-chatbot environments. We have already implemented this approach, but have not been able to fully test it.
Another social limitation concerns the fact that the bot itself does not recruit survey respondents, which is currently the biggest challenge for social and marketing research.

\section{Conclusion}
\emph{MyMigrationBot} is the first deployment of a Facebook social chatbot using Twilio in Central Europe, and the second in Europe~\cite{chen2020creating} to support migrant population. To our knowledge, it is also the first chatbot using Facebook's Graph API for information retrieval, to gather reliable Big Data for social science research, with a special focus on migrant populations.

The application was positively assessed by our beta testers, who were both migrants and stayers.
There was no significant statistical difference in perceived usability between the groups. People do not like filling out surveys but can be motivated by constructive feedback given to them~\cite{kuhne2018personalized}. We have gathered a lot of interesting feedback from 53 beta testers. We will focus our efforts on improving user experience e.g. by adding graphical elements to the surveys, and overcoming the key limitations highlighted by our users in beta testing, prior to publicly releasing the \emph{MyMigrationBot} for migrant populations. We also plan to place our MyMigrationBot into a co-design process with migrants in focused groups~\cite{chen2020creating} in order to: firstly, maximise an interactive engagement of our users and therefore a conversation flow; secondly, to enhance the usability of the machine; and thirdly, to refine our initially proposed solution. We also plan to deliver our \emph{MyMigrationBot} to migrant population as part of a platform for assessments of various psycho-social capitals of migrants which are used in the labour market.

\section*{Acknowledgment}
We thank Jennifer Fei and Jessica Sadye Wolff from Immigration Policy Lab at Stanford University and Michał Kosiński from Stanford Graduate Business School for several discussions that led to the development of~\emph{MyMigrationBot}. We thank Michael Connolly for proofreading, Dominika Winogrodzka and Ivanna Kyliushyk for their on-going support with \emph{MyMigrationBot}. The publication was supported by a grant from National Science Centre (Poland) no. 2020/37/B/HS6/02342.

\printbibliography
\section*{Appendix}
\begin{table*}[tbp]
\caption{{MyMigrationBot}'s logical data model summary}\label{RDSDatabaseStructure}
\centering
\begin{tabular}{|@{\vrule width0ptheight9pt\enspace}l|c|c|c|c|c|c|}\hline
\hfil\bf Attribute Name&\bf Type& \bf Default&\bf Source&\bf Description\\\hline

\verb|Id (PRIMARY KEY)|&bigint& NOT NULL&Database & Primary key of the record in database \\\hline
\verb|Fb_Id|&varchar(255)& NOT NULL&Messenger protocol& Facebook user identifier \\\hline
\verb|First_name|&varchar(255)&NULL&Graph API &  FB user's first name\\\hline
\verb|Last_name|&varchar(255)& NULL&Graph API& FB user's last name \\\hline
\verb|Locale|&varchar(255)& NULL&Graph API & FB user's home address \\\hline
\verb|Hometown|&varchar(255)& NULL&Graph API & FB user's birth place\\\hline
\verb|Timezone|&float& NULL& Graph API&FB user's current timezone offset from UTC\\\hline
\verb|Birthday|&varchar(255)& NULL&Graph API &FB user's birthday\\\hline
\verb|Gender|&varchar(255)& NULL&Graph API &FB user's gender\\\hline
\verb|TIPIPL_odp_1|&smallint& NULL&User input&User answer for TIPIPL question \\\hline
\verb|TIPIPL_odp_2|&smallint& NULL&User input& User's answer for TIPIPL question \\\hline
\verb|TIPIPL_odp_3|&smallint& NULL&User input& User's answer for TIPIPL question \\\hline
\verb|TIPIPL_odp_4|&smallint& NULL&User input& User's answer for TIPIPL question \\\hline
\verb|TIPIPL_odp_5|&smallint& NULL&User input& User's answer for TIPIPL question \\\hline
\verb|TIPIPL_odp_6|&smallint& NULL&User input&User's answer for TIPIPL question \\\hline
\verb|TIPIPL_odp_7|&smallint& NULL&User input&User's answer for TIPIPL question \\\hline
\verb|TIPIPL_odp_8|&smallint& NULL&User input&User's answer for TIPIPL question \\\hline
\verb|TIPIPL_odp_9|&smallint& NULL&User input&User's answer for TIPIPL question \\\hline
\verb|TIPIPL_odp_10|&smallint& NULL&User input&User's answer for TIPIPL question \\\hline
\verb|TIPIPL_ekstarwersja|&double(5,1)& NULL&Backend&Evaluation of user's answer for Extraversion \\\hline
\verb|TIPIPL_ugodowosc|&double(5,1)&  NULL&Backend&Evaluation of user's answer for Agreeableness \\\hline
\verb|TIPIPL_sumiennosc|&double(5,1)& NULL&Backend&Evaluation of user's answer for Conscientiousness \\\hline
\verb|TIPIPL_stabilnosc|&double(5,1)& NULL&Backend&Evaluation of user's answer for Emotional Stability \\\hline
\verb|TIPIPL_otwartosc|&double(5,1)& NULL&Backend&Evaluation of user's answer for Openness to Experience \\\hline
\verb|DopKomp_czy_pracujesz|&varchar(5)& NULL&User input&Are you employed? (Flag)\\\hline
\verb|DopKomp_odp_num_1|&varchar(5)& NULL&User input&User's answer to Competence-Job fit question\\\hline
\verb|Inter_odp_1|&smallint& NULL&User input&User's answer to System Usability Scale\\\hline
\verb|Inter_odp_2|&smallint& NULL&User input&User's answer to System Usability Scale\\\hline
\verb|Inter_odp_2|&smallint& NULL&User input&User's answer to System Usability Scale\\\hline
\verb|Inter_odp_3|&smallint& NULL&User input&User's answer to System Usability Scale\\\hline
\verb|Inter_odp_4|&smallint& NULL&User input&User's answer to System Usability Scale\\\hline
\verb|Inter_odp_5|&smallint& NULL&User input&User's answer to System Usability Scale\\\hline
\verb|Inter_odp_6|&smallint& NULL&User input&User's answer to System Usability Scale\\\hline
\verb|Inter_odp_7|&smallint& NULL&User input&User's answer to System Usability Scale\\\hline
\verb|Inter_odp_8|&smallint& NULL&User input&User's answer to System Usability Scale\\\hline
\verb|Inter_odp_9|&smallint& NULL&User input&User's answer to System Usability Scale\\\hline
\verb|Inter_odp_10|&smallint& NULL&User input&User's answer to System Usability Scale\\\hline
\verb|Record_created|&datetime& NULL&Backend&Database date and time of record creation\\\hline
\verb|Jezyk| &varchar(10)& NULL&User input&Language flag for the surveys executed\\\hline
\verb|Profile_pic| &varchar(500)& NULL&Graph API&User's Facebook profile picture (Url) \\\hline
\verb|Age| &smallint& NULL&Graph API&Age of Facebook user\\\hline
\verb|It_skils| &smallint& NULL&Graph API&FB user's level of it skils (1-5)\\\hline
\verb|Immigrant| &smallint& NULL&Graph API& FB user's immigrant flag\\\hline
\verb|Device| &varchar(500)& NULL&Graph API& FB user's device type flag\\\hline
\end{tabular}
\end{table*}

\begin{table*}[tbp]
\caption{Participants' feedback on MyMigrationBot}\label{ParticipantFeedback}
\centering
\begin{tabular}{|@{\vrule width0ptheight9pt\enspace}l|c|}\hline
\hfil\bf What are your overall impressions of using this chatbot on Facebook?\\\hline
- \say{Generally a good impression. Easy to use, readable.}\\
- \say{Chatbot was ok to use, the questions were easy to understand. There were no technical problems.}\\
- \say{It was fun, the questions were interesting. It gave me something to think about.}\\
- \say{Easy to use, horrible in visual design.}\\
- \say{Pretty good, sometimes the languages were not consistent}\\
- \say{Quite short, and not a bad experience overall.}\\
- \say{When I inputted a random letter, the chatbot corrected me – asked to use one of 5 valid answers, which is nice to see.}\\
- \say{The program is friendly and convenient, the survey can be completed anywhere, and only takes a short while.}\\
- \say{At first I did not know how to start, nothing was displayed until I made the first move. I had to initiate the conversation.}\\
- \say{I didn't know what to do, when the greeting in three languages came out.}\\
- \say{It gives a person a feeling of communication with a person, and therefore some comfort.}\\\hline
 \hfil\bf What are the advantages of the chatbot experience/interaction?\\\hline
    - \say{Easy to understand questions, variety of languages.}\\
    - \say{It was fun to look into ourselves and to see how we are.} \\
    - \say{The questions were interesting and it gave me a bit to think about myself and how I behave around people.}\\
    - \say{Not having to scroll through different pages and check boxes, using natural language, even if it was just typing in numbers, was a lot more natural.}\\
    - \say{Easy to use. Fast. Easy to use/read, few languages.It’s easily  accessible.}\\
    - \say{It was fast, very simple to use, pretty clear and non-sophisticated.}\\ 
    - \say{Also the feedback (input only values 1-5) in case of a wrong answer was a nice thing.}\\
    - \say{It was nice to see in what category I belong to. It helps us have a better idea of who we are. It wasn’t too long to take.}\\\hline
     \hfil\bf What are the disadvantages of the chatbot experience/interaction?\\\hline
    - \say{Different scale for each question (sometimes 1-7 or 1-5,1-6) which is confusing and easy to miss, also I do not find this bot helpful, it is not complex enough.}\\
    - \say{When I chose English language, there are some sentences in Polish.}\\
    - \say{Scales across all surveys should be the same. At the moment the psychological tools use the scale from 1-5, 1-6 or from 1 to 7.}\\
    - \say{Executing survey in English, I found a few chatbot answer in Polish.}\\
    - \say{I didn't know how to start the conversation with the social chatbot.}\\
    - \say{There is No ability to change answers.}\\
    - \say{Lack of the option of choosing the answer by clicking the mouse, currently they must type the answer from the keyboard.}\\
    - \say{It’s a little more difficult to read through the longer messages.}\\
    - \say{The bot should print the messages in order. Sometimes it sent multiple different messages in the same time (results + questions).}\\
     - \say{The worst was the fact that in the end there were questions (are you employed) which were put in between other results statements which was very confusing.}\\
     - \say{User should not be given questions and statements in random order.}\\
    - \say{Regrading adding answers, using keyboard is less convenient that clicking on the links in some kind of form.}\\
    - \say{After selecting English as a language – some of the messages were still written in Polish.}\\
    - \say{There was also no clear instruction in the chatbot itself to begin the conversation.}\\\hline
     \hfil\bf What made it difficult for you to use/interact with the chatbot?\\\hline
    - \say{Sometimes it was hard put a number on any expression. Other than that I had no difficulties.}\\
    - \say{Really easy to use if you have eyes to read. It's hard to read long messages on the messenger.}\\
    - \say{There were some hiccups with the language the bot used. Even though I selected English at the beginning, parts of the messages were in Polish it was confusing.}\\
    - \say{Nothing, I was only afraid of typing something wrong – I wouldn’t like to break the chatbot.}\\ 
    - \say{Finding questions in such long pieces of text was a bit challenging.}\\
    - \say{In the end the question was combined with a “Thank you for completing….etc.” and I didn’t see there was a question after all.}\\

    - \say{Lack of instructions Inconsistency in language Messages appearing too quickly.I didn’t know if I should read or answer the next ones.}\\
    - \say{The scale (1-7;1-5) was very complicated. In chatbots without visuals maybe it would be easier to have simpler scale like 1-3.} \\\hline
     \hfil\bf How can we improve the experience/interaction with the chatbot?\\\hline
    - \say{I would also recommend to add a 'sleep time' for the bot (wait few seconds before the bot sends its response).}\\
    - \say{In the questions maybe disclose different behaviours, but similar, so that a person while using the chatbot would not feel that they have to choose only one.}\\
    - \say{Instead create more questions that might give you more insight to the person.}\\
    - \say{I would recommend to add “sleep time” for the bot (wait few seconds before boot sends its respond).}\\
    - \say{Any sign that bot is working, gives a person feeling of communication with a person and therefrom – comfort.}\\
    - \say{Split questions for sure. Some different traits were put together and while I felt 3 with one, I felt 7 with other. So I chose more or less 5 but that wasn’t good.}\\
    - \say{In the final part no matter what language you choose you get a sum-up in polish which may be off-putting for some people because they wouldn’t understand it.}\\
    - \say{Maybe also split the long blocks of text into smaller ones – especially if there are questions mixed with statements.}\\
    - \say{Instead of having the user type in numbers, use chatbot buttons.}\\
    - \say{Maybe at the end, the chatbot can compile a small file of all the results to send to the user.}\\
    - \say{shorter messages. It could be anonymous.}\\
    - \say{Make buttons to click on, instead of having numbers to type. Reduce the repetitive questions.}\\
    - \say{Unify the language after choosing one.}\\
    - \say{Add intro like: Hi I am a chatbot that will help you to.}\\\hline
\end{tabular}
\end{table*}

\end{document}